\documentclass[aps,floatfix,twocolumn]{revtex4}
\usepackage{amssymb,amsmath,epsf}
\usepackage[dvips]{color}
\usepackage{framed,multirow}

\usepackage{graphicx}

\begin{document}
\title{Real-space grid representation of momentum and kinetic energy
operators for electronic structure calculations}
\author{Domenico Ninno}
\author{Giovanni Cantele}
\affiliation{CNR-SPIN Napoli and Universita' degli Studi di Napoli Federico II,
Dipartimento di Fisica Ettore Pancini, Complesso Universitario di Monte S. Angelo,
Via Cintia, I-80126 Napoli, Italy.}
\author{Fabio Trani}
\email{fabio.trani@unina.it}
\affiliation{Universita' degli Studi di Napoli Federico II,
Dipartimento di Fisica Ettore Pancini, Complesso Universitario di Monte S. Angelo,
Via Cintia, I-80126 Napoli, Italy.}
\begin{abstract}
We show that the central finite difference formula for the first and the
second derivaxtive of a function can be derived, in the context of quantum
mechanics, as matrix elements of the momentum and kinetic energy operators
using, as a basis set, the discrete coordinate eigenkets 
$\vert x_n\rangle$ defined on the uniform grid $x_n=na$. Simple closed form
expressions of the matrix elements are obtained starting from integrals
involving the canonical commutation rule.
%$\left[\widehat{x},\widehat{p}\right]=i\hbar\widehat{I}$. 
A detailed analysis of the convergence toward 
the continuum limit with respect to both the grid spacing 
and the approximation order is presented. It is shown that the 
convergence from below of the eigenvalues in electronic structure
calculations is an intrinsic feature of the finite difference method. 
\end{abstract}

\maketitle
%\begin{keyword}
%\PACS 31.15.xf 71.15.Ap 71.15.Dx
%\end{keyword}

%%%%%%%%%%%%%%%%%%%%%%%%%%%%
\section{Introduction}
\label{intro}
%%%%%%%%%%%%%%%%%%%%%%%%%%%%
The last decade has witnessed a growing interest in electronic structure
computational schemes\cite{collection,beck,gpaw,rescu,hernandez} based on the real-space finite difference 
representation of both the momentum and kinetic energy operators.
The main advantage of the method when 
applied to either the Schr\"{o}dinger or the DFT
Kohn and Sham equations\cite{kohn}, particularly within 
the pseudopotential density functional theory\cite{cheli1}, is
the highly local structure of the relevant
matrices. This feature allows for both a significant
reduction of computer memory and for the use of efficient parallel algorithms
\cite{burdick} paving the way toward grid-based linear scaling methods. 
Ab initio electronic structure\cite{beck,cheli1,fattebert,octopus,andrade15}
and molecular dynamics codes\cite{md04}
have successfully been implemented and tested. Moreover,
adaptive mesh refinements\cite{modine} and multigrid strategies\cite{briggs}
can be used for gaining local resolution only where it is needed. 
Finally, it is worth mentioning that the finite difference method is
also a very good conceptual\cite{datta} 
and practical\cite{fujimoto,khomyakov} tool
for the study of electric current flow in nanodevices. 

Other electronic structure computational methods closely related to real
space grids are finite elements schemes\cite{white}, discrete variable
representations\cite{tuckerman}, Lagrange meshes\cite{varga}, nonorthogonal
generalized Wannier functions\cite{skylaris2},
and wavelets\cite{arias,genovese}. All
these methods have in common a real space grid whose points are associated
to localized basis functions\cite{baye}. As such, it seems that there is not connection
between this class of methods and finite differences.  We shall show in
the following that indeed there is a link when the continuum limit
in the finite differences is approached. 

For entering the core of the problem discussed here, let us
briefly recall how the finite difference method works.
The starting point is the choice of the way
in which the kinetic energy operator is represented 
on a discretized space. Let us consider a one
dimensional grid (the extension to two and three dimensional Cartesian grids
is trivial) made of $N+1$ equally spaced points defined by $x_{n}=na$ 
with $n=0,1..,N$. The lowest-order approximation to the second derivative is
obtained considering only three grid points $x_{n+1},$ $x_{n}$ and $x_{n-1}$.
With a Taylor series it is easy to see that the lowest order approximation
of second derivative at $x_n$ is given by
\begin{equation}
\frac{d^{2}\psi(x_n)}{dx^{2}}\approx\frac{1}{a^{2}}\left[ \psi(x_{n-1})-2\psi
(x_{n})+\psi(x_{n+1})\right]  \label{eq0}.
\end{equation}
Higher order expressions involving wave function values at $x_{n\pm m}$ can
be obtained by means of specific algorithms\cite{fornberg}. In general, a
finite difference representation of the second derivative takes the form 
\begin{equation}
\frac{d^{2}\psi(x_n)}{dx^{2}}\approx\sum_{m=-M}^{M}C_{m}\psi(x_{n+m}) 
\label{eq00},
\end{equation}
where the integer number $M$, which we 
define as the representation order,  controls the accuracy.  
Whichever method is used for determining the coefficients $C_{m}$, it is
clear that the kinetic energy representation induced by  
Eq. (\ref{eq00}) does 
not apparently have an explicit basis set.
In other words, although the $C_{m}$ can formally be viewed as the entries
of the kinetic energy matrix, it is not at all clear 
whether or not there exists an underlying basis
set. This is, in our opinion, an interesting and fertile point that has not been
discussed in the literature. 

A further point is represented by the convergence properties 
toward the continuous space, that is, the limit 
of vanishing grid spacing. 
For instance, practical finite difference 
DFT electronic structure calculations show that
the total energy converges from below with respect to both the grid spacing and 
the order $M$\cite{beck,kaxiras,skylaris1}.  It is generally
believed that this is an important limitation preventing from developing
convergence schemes. 

The core of this work is first of all to show that the finite difference
method has discrete coordinate eigenkets as the underlying basis set. 
We obtain
this result by explicitly constructing  the matrix elements of both the linear
momentum and kinetic energy operators. 
Interestingly, the unique starting
point of the derivation is the use of integrals 
involving the canonical commutations
between the operators $\widehat{x}$ and $\widehat{p}$. 
In a way, we leave to the quantum mechanics rules the determination
of the optimal matrix elements.  
We show that it is possible to obtain simple analytical
expressions for the matrix elements so that their convergence towards
the continuum limit and the connection with other real space methods 
can be discussed in detail. The theory is complemented 
with some explanatory numerical examples.
%%%%%%%%%%%%%%%%%%%%%%%%%%%
\section{The momentum operator}
\label{sec1}
%%%%%%%%%%%%%%%%%%%%%%%%%%%
Let us associate to each grid point $x_n$ the ket 
$\vert x_{n}\rangle$ which we assume to be
an eigenket of the position operator $\hat{x}$. This means that 
\begin{equation}
\langle x_{m}\vert\hat{x}\vert x_{n}\rangle =
x_{n} \delta _{m,n}.
\label{exp2.1}
\end{equation}
Any state defined
on this discretized space  can be written as 
\begin{equation}
\left\vert \psi\right\rangle =\sum_{n}\phi_{n}\left\vert x_{n}\right\rangle .
\label{exp1}
\end{equation}
Although it is not strictly necessary for the present discussion, 
we can notice here that the numbers
$\phi_{n}$ in Eq. (\ref{exp1}) are the values of the wave function $ \psi(x)$
on the grid points $x_n$.  

A grid representation of the momentum operator is given by 
\begin{equation}
\left\langle x_{n}\right\vert \hat{p}\left\vert f\right\rangle
=-i\hbar\sum_{m=-M}^{M}W_{m}\left\langle x_{n+m}\right. \left\vert
f\right\rangle   \label{eq1},
\end{equation}
where $\left\vert f\right\rangle $ is a generic ket. 
The real numbers $W_{m}$ are the 
momentum matrix elements. 
For a given $M$
there are $2M+1$ grid points over which the $W_{m}$ are different from zero.
This means that the set of $W_{m}$ depends on $M$. For indicating explicitly
this dependence, we write, from now on, $W_{m}^{(M)}$ and
call $M$ the representation order. Taking $\left\vert f\right\rangle
=\left\vert x_{l}\right\rangle $, from Eq. (\ref{eq1}) we have 
\begin{equation}
\left\langle x_{n}\right\vert \hat{p}\left\vert x_{l}\right\rangle
=-i\hbar\sum_{m=-M}^{M}W_{m}^{(M)}\delta_{n+m,l}=-i\hbar W_{l-n}^{(M)},
\label{eq3}
\end{equation}
where $-M\leq l-n\leq M$. 
Since we want an Hermitian matrix representing $\hat{p}$, we must require 
\begin{equation}
W_{-m}^{(M)}=-W_{m}^{(M)}  \label{cond1},
\end{equation}
and this implies that $W_{0}^{(M)}=0$. 
Additional conditions necessary for determining 
$W_{m}^{(M)}$ can be derived
starting from integrals
involving the canonical commutation
\begin{equation}
\left[ \hat{x},\hat{p}\right] =i\hbar\hat{I}  \label{eq5}.
\end{equation}
With continuum position eigenkets it is known that 
\begin{equation}
\int\left\langle x^{\prime}\right\vert \left[ \hat{x},\hat{p}\right]
\left\vert x^{\prime\prime}\right\rangle dx^{\prime\prime}=i\hbar .
\label{cont2}
\end{equation}
The discrete analog of Eq. (\ref{cont2}) is obtained by calculating the matrix
elements $\langle x_{n}\vert \left[ \hat{x},\hat {p}
\right] \vert x_{n+m}\rangle$ and summing on $m$. Using Eq. (\ref{eq3}) we have
\begin{equation}
\sum_{m=-M}^{M}\left\langle x_{n}\right\vert \left[ \hat{x},\hat {p}%
\right] \left\vert x_{n+m}\right\rangle =i\hbar a\sum_{m=-M}^{M}mW_{m}^{(M)} .
\label{eq6}
\end{equation}
It follows that if we want that Eq. (\ref{eq6}) gives exactly the same result as
Eq. (\ref{cont2}), the matrix elements 
$W_{m}^{(M)}$ must satisfy the equation 
\begin{equation}
\sum_{m=1}^{M}mW_{m}^{(M)}=\frac{1}{2a},
\label{cond2}
\end{equation}
where we have used Eq. (\ref{cond1}) for limiting the sum to positive 
$m$. Equation
(\ref{cond2}) is just enough for determining $W_{m}^{(M)}$ with $%
M=1$. Indeed, one immediately gets $W_{\pm 1}^{(1)}=\pm 1/2a$ 
which are exactly
the central difference weights for the lowest order first derivative. 
This an interesting intermediate result. However, the issue now is to show how
the most general case of $M>1$ can be handled, that is, how we can generate
the necessary additional equations. Since multiple commutators
of the type $\left[ \hat{x},\left[ \hat {x},
\ldots, \left[ \hat{x}
,\hat{p}\right] \right]\right]$ are all zero, we have
\begin{equation}
\int\left\langle x^{\prime}\right\vert \left[ \hat{x},\left[ \hat {x},
\ldots, \left[ \hat{x}
,\hat{p}\right] \right]\right] \left\vert x^{\prime\prime}\right\rangle
dx^{\prime\prime}=0.
\label{comm2}
\end{equation}
In analogy with Eqs. (\ref{cont2}) and (\ref{eq6}), the discretization of this equation leads us to
\begin{equation}
\sum_{m=-M}^{M}\left\langle x_{n}\right\vert \left[\hat{x},\left[ 
\hat{x},\ldots,\left[ \hat{x},\hat{p}\right] \right]\right] \left\vert x_{n+m}\right\rangle
=\sum_{m=-M}^{M}m^{s}W_{m}^{(M)}=0   
\label{comm3}
\end{equation}
where $s$ is the number of times $\hat{x}$ appears in the multiple commutator.
Using Eq. (\ref{cond1}) it is easy to see that Eq. (\ref{comm3}) is identically
satisfied when $s$ is even. We are therefore left with
\begin{equation}
\sum_{m=1}^{M}m^{2l+1}W_{m}^{(M)}=0%
\text{ where }l=1,\ldots M-1   
\label{cond3}.
\end{equation}
For a given $M$, the set of Eqs. (\ref{cond3})
and (\ref{cond2}) can be solved. The final result is 
\begin{equation}
W_{m}^{(M)}=\frac{1}{2am\Omega(M,m)}  ,
\label{final1}
\end{equation}
where we have defined 
\begin{equation}
\Omega(M,m)=\prod\limits_{\substack{l=1\\l\neq m}}^{M}
\left[1-\left(  \frac{m}{l}\right)^{2}\right]  .
\label{final2}%
\end{equation}
The momentum matrix elements are
\begin{equation}
\left\langle x_{n}\right\vert \hat{p}\left\vert x_{n+m}\right\rangle
=\left\{
\begin{array}
[c]{ll}%
0 & \text{if }m=0\\
\displaystyle{\frac{-i\hbar}{2am\Omega\left(  M,m\right) } } & \text{if }m\neq0
\end{array}
\right.
\label{eqmom}
\end{equation}
The numerical values of Eq. (\ref{eqmom}) are identical to 
those obtained for the first derivative central finite
difference weights using the algorithm discussed in Ref.\cite{fornberg}. 
It is useful for the following to calculate the limit of Eq. (\ref
{final2}) when $M\rightarrow \infty$. The result is extraordinarily simple
\begin{equation}
\Omega(\infty,m)=\frac{\left( -1\right) ^{m+1}}{2}   \label{final3},
\end{equation}
from which we immediately have 
\begin{equation}
\left\langle x_{n}\right\vert \hat{p}\left\vert x_{n+m}\right\rangle
=\left\{
\begin{array}
[c]{ll}%
0 & \text{if }m=0\\
\displaystyle{\frac{i\hbar (-1)^{m}}{am } } & \text{if }m\neq0
\end{array}
\right.
\label{mominfty}
\end{equation}
The actual meaning and implications of this result 
will be discussed in the next section.

Before closing this section, it may be useful to summarize
the main points. The quantum mechanical 'rules' expressed in 
Eq. (\ref{cont2}) and (\ref{comm2}) have guided us 
to the simple closed form expression (\ref{final1})
of the weights $W_{m}^{(M)}$. Since our derivation has
been built on the basis set $\vert x_n\rangle$, we have
indirectly proved that this is the basis set of the finite difference scheme.    
%%%%%%%%%%%%%%%%%%%%%%%%%%%%%%%
\section{The kinetic energy operator}
\label{sec2}
%%%%%%%%%%%%%%%%%%%%%%%%%%%%%%%
We now derive an expression for the kinetic energy matrix elements following 
steps that are similar to what we have done 
for the linear momentum. The kinetic energy can be defined as 
\begin{equation}
\left\langle x_{n}\right\vert \hat{T}\left\vert f\right\rangle =-\frac{%
\hbar ^{2}}{2\mu }\sum_{\substack{ l=-M \\ l\neq 0}}^{M}C_{l}^{(M)}\left(
\left\langle x_{n+l}\right. \left\vert f\right\rangle -\left\langle
x_{n}\right. \left\vert f\right\rangle \right)   \label{cin1}.
\end{equation}%
From Eq. (\ref{cin1}) and the orthogonality of 
the set $\vert x_n \rangle$ it is easy to see that
\begin{equation}
\langle x_{n}|\hat{T}|x_{n+m}\rangle=\frac{\hbar^{2}}{2\mu}\left\{
\begin{array}
[c]{ll}%
-C_{m}^{(M)} & \text{ if }m\neq0 \vspace{0.2cm}\\
\sum\limits_{\substack{l=-M\\l\neq0}}^{M}C_{l}^{(M)}& \text{ if }m=0
\end{array}
\right.  \label{cin2}%
\end{equation}

The coefficients $C_m^{(M)}$ in Eq. (\ref{cin2}) can be related to $W_m^{(M)}$ taking the matrix elements of
$\hat{p}=\frac{\mu}{i\hbar}\left[\hat{x},\hat{T}\right] $  
on the basis $\vert x_n \rangle$. The final result is
\begin{equation}
C_m^{(M)}=\frac{2W_m^{(M)}}{am}
\label{cin4}.
\end{equation}
Inserting Eq. (\ref{cin4}) into Eq. (\ref{cin2}) we finally have
\begin{equation}
\left\langle x_{n}\right\vert \hat{T}\left\vert x_{n+m}\right\rangle
=\frac{\hbar^{2}}{2\mu a^{2}}\left\{
\begin{array}
[c]{ll}%
\displaystyle{\frac{-1}{m^{2}\Omega(M,m)}} & \text{ if }m\neq0 \vspace{0.2cm}\\
\displaystyle{\sum_{l=1}^{M}\frac{2}{l^{2}\Omega(M,l)}} & \text{ if }m=0
\end{array}
\right.
\label{cin5a}
\end{equation}
Eq. (\ref{cin5a}) gives a closed form expression for the kinetic energy 
valid to all orders $M$. 
The advantage of having an analytical expression of the matrix elements 
is not only the simplicity with which they can be generated in an
electronic structure computer code, but also the possibility of easily looking
at their behavior when the representation order $M$ is very large. Indeed,
from Eq. (\ref{final3}) it is easy to see that when $M\rightarrow\infty$    
Eq. (\ref{cin5a}) becomes
\begin{equation}
\left\langle x_{n}\right\vert \hat{T}\left\vert x_{n+m}\right\rangle
=\frac{\hbar^{2}}{2\mu a^{2}}\left\{
\begin{array}
[c]{ll}%
\displaystyle{\frac{2(-1)^{m}}{m^{2}}} & \text{ if }m\neq0 \vspace{0.2cm}\\
\displaystyle{\frac{\pi^{2}}{3}} & \text{ if }m=0
\end{array}
\right.  
\label{cininfty}
\end{equation}

It should be noted that although Eq. (\ref{cininfty}) is strictly valid 
only for a grid made of an infinite 
number of points, it  is nevertheless an interesting results because 
it establishes a connection between
finite difference and pseudospectral methods\cite{mazz1,baye}. It is worth recalling that in
a pseudospectral method the wave function is written as\cite{baye}
\begin{equation}
\psi(x)=\sum_n a_n C_n(x)
\label{ps1},
\end{equation}
where the basis $C_n(x)$ has the cardinal property
\begin{equation}
C_n(x_m)=\delta_{m,n}.
\label{ps2}
\end{equation}
Because of this property, Eq. (\ref{ps1}) becomes
\begin{equation}
\psi(x)=\sum_n \psi(x_n) C_n(x).
\label{ps3}
\end{equation} 
Although there is a similarity between Eqs. (\ref{ps3}) and (\ref{exp1}),
it should be observed that while in the former the basis 
and the wave function are  defined in a
continuous space, in the latter they are
defined only on the grid points. In a way we can say that Eq. (\ref{ps3}) is
the continuum space extension of Eq. (\ref{exp1}). 

Let us now calculate the 
matrix elements of the kinetic energy 
operator $\hat{T}=-\frac{\hbar^2}{2\mu}\frac{d^2}{dx^2}$ 
on the $sinc$ (also known as Whittaker cardinal)
functions
\begin{equation}
sinc(x-x_n)=\frac{sin\left(\pi(x-x_n)/a\right)}{\pi(x-x_n)/a}.
\label{sinc}
\end{equation}     
It is quite interesting that the result of 
this calculation\cite{mazz1,baye} is exactly Eq. (\ref{cininfty}).
In a similar way, it is possible 
to show that the momentum matrix elements 
between $sinc$ functions are 
given by Eq. (\ref{mominfty}). 
We therefore see that
the connection between the finite difference and the pseudospectral methods
goes through the infinite order limit $M\rightarrow\infty$.

%%%%%%%%%%%%%%%%%%%%%%%%%%%%%%%%%
\section{The convergence towards the continuous space}
\label{sec3}
%%%%%%%%%%%%%%%%%%%%%%%%%%%%%%%%%
For having more insights on the convergence properties
with respect to $M$
it is necessary to give a sense to 
the limit $M\rightarrow\infty$ on a space grid
made of a finite number of points. As it is well known in 
solid state theory, an effective way of constructing an
infinite lattice out of a finite one is that of taking periodic replicas.
We can use exactly the same idea by taking periodic replicas 
of a grid of length $L=Na$. In this way  
the limit $M\rightarrow\infty$ has a precise meaning.

On a replicated grid we need to impose periodic boundary conditions.
We shall see in a moment that this choice is particularly useful
because it allows the analytical calculation 
of both the momentum and the kinetic energy
eigenvalues. 
Let us show this for the momentum matrix. 
The eigenvalue equation is 
\begin{equation}
\sum_{m=-M}^{M}\langle x_n \vert \hat{p}\vert x_{n+m}\rangle
\left\langle x_{n+m}\right. \left\vert
k\right\rangle =p\left\langle x_{n}\right. \left\vert k\right\rangle 
\label{mom5},
\end{equation}
where we have indicated with $\langle x_n\vert k\rangle$ the momentum
eigenvectors on the grid. It is a simple matter to verify that
Eq. (\ref{mom5}) can be solved with
\begin{equation}
\left\langle x_{n}\right. \left\vert k\right\rangle =e^{-ikx_{n}} 
\label{mom6}.
\end{equation}
The wave vector $k$ is selected imposing 
the periodic boundary condition $\left\langle x_{n+N}\right.
\left\vert k\right\rangle =\left\langle x_{n}\right. \left\vert
k\right\rangle $ giving the well known result
\begin{equation}  
k_{\nu}=\frac{2\pi\nu}{L} \text{ with } \nu=0,1,2\ldots N-1
\label{kselect}
\end{equation}
where $L=Na$ is the grid length.  
Inserting Eq. (\ref{mom6}) into Eq. (\ref{mom5}) 
and using Eq. (\ref{final1}) we have the eigenvalues
\begin{equation}
p_{\nu}=\frac{\hbar}{a}\sum_{m=1}^{M}\frac {%
\sin\left( x_m k_{\nu}\right) }{m\Omega(M,m)}   \label{mom7}.
\end{equation}
The simple structure of Eq. (\ref{mom7}) allows us to
study in detail the convergence towards the continuum with respect to both
the representation order $M$ and the grid spacing $a$. 
%%%%%%%%%%%%%%%%%%%%%%%%%%%%%
%FIGURE 1
\begin{figure} [t]
\includegraphics[scale=0.3]{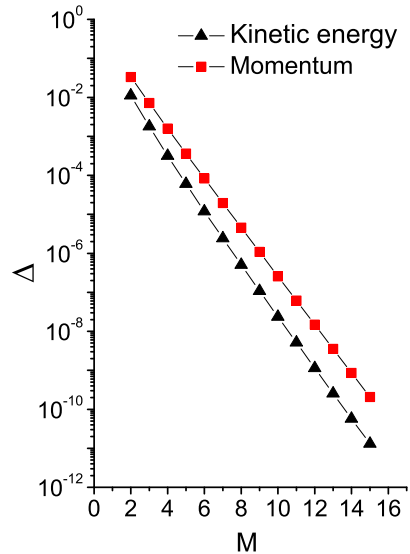}
\caption{(Color online) The correction $\Delta$ as defined in Eq. (\ref{eq36})
for the momentum and Eq. (\ref{eq39}) for the kinetic energy as a 
function of $M$.}
\label{fig1}
\end{figure}
%FIGURE 1
%%%%%%%%%%%%%%%%%%%%%%%%%%%%%%  
However, before doing that, let us also see what we obtain for the kinetic energy.
From an eigenvalue equation similar to Eq. (\ref{mom5}) we get
\begin{equation}
\varepsilon_{\nu}=\frac{\hbar^{2}}{2\mu a^{2}}\sum_{m=1}^{M}
\frac{4}
{m^{2}\Omega(M,m)} \sin^{2}\left(\frac{ x_m k_{\nu}}{2}\right ) 
\label{kin1}.
\end{equation}

We are now ready to take the limit $M\rightarrow \infty$. 
For the momentum we must use Eqs. (\ref{mom7}) and (\ref{final3}). 
The result is
\begin{equation}
p_{\nu}=\frac{\hbar}{a}\sum_{m=1}^{\infty}\frac{2\left( -1\right) ^{m+1}}{m}%
\sin\left(x_m k_{\nu}\right) =\hbar k_{\nu}   .
\label{mom3}
\end{equation}
This is quite interesting: it tell us that with 
a fixed grid spacing $a$, the linear 
momentum gets exactly the values one would expect 
for a free particle on a lattice of length $L=Na$ with
periodic boundary conditions.
The type of convergence behind Eq. (\ref{mom3})
can be better appreciated by computing 
the power series of Eq. (\ref{mom7}) with respect to
the grid spacing $a$. A lengthy calculation in which we 
retain, for each order $M$, the
first non vanishing terms gives
\begin{equation}
p_{\nu}=\hbar k_{\nu}\left[  1-\left( k_{\nu}a \right)  ^{2M}\Delta_{p}(M)+\ldots\right]
\label{eq35}
\end{equation}
where 
\begin{equation}
\Delta_{p}(M)=\frac{2(M+1)}{\left(  2M+2\right)  !}\left\vert \sum_{m=1}%
^{M}\frac{m^{2M}}{\Omega (M,m)}\right\vert
\label{eq36}.
\end{equation}
It should be noted that since $\Delta_{p}(M)$ is always positive, 
Eq. (\ref{eq35}) shows that the convergence to the 
exact momentum is from below. 

The limit $M\rightarrow \infty$ for the kinetic energy is easily obtained from
Eqs. (\ref{kin1}) and (\ref{final3}). The result is
\begin{equation}
\varepsilon_{\nu}=\frac{\hbar^{2}}{2\mu a^2}
\sum_{m=1}^{\infty}\frac{4\left( -1\right)
^{m+1}}{m^{2}}\sin^{2}\left( x_mk_{\nu}\right) =
\frac{\hbar^{2}k_{\nu}^2}{2\mu }%
  \label{eq37}.
\end{equation}
Again, we get the exact kinetic energy of a free particle on a lattice
with periodic boundary conditions..
In analogy with Eq. (\ref{eq35}), a power series of Eq. (\ref{kin1})
leads to
\begin{equation}
\varepsilon_{\nu}=\frac{\hbar^2k_{\nu}^2}{2\mu}
\left[  1-\left( k_{\nu}a\right)  ^{2M}\Delta_{\varepsilon}(M)\ldots\right]
\label{eq38},
\end{equation}
where we have defined
\begin{equation}
\Delta_{\epsilon}(M)=\frac{2}{\left(  2M+2\right)  !}\left\vert \sum_{m=1}%
^{M}\frac{m^{2M}}{\Omega (M,m)}\right\vert
\label{eq39}.
\end{equation}
The same conclusions
 we have drawn for the momentum hold in this case.
In particular, it is worth stressing the convergence from below which plays
an important role in electronic structure calculations. 

To give a hint on
how small the 
the relative error $\Delta$ are, we show 
in Fig. \ref{fig1} a plot of Eqs. (\ref{eq36}) and (\ref{eq39}) as a function of $M$. 
It can be seen from this figure that $\Delta$ quickly goes 
down so that one would expect that in a practical
numerical calculation there should be no need of using values of $M$
larger than about $6\div8$. This is important because  
keeping a banded kinetic energy matrix is a prerequisite 
for a fast calculation. 
%%%%%%%%%%%%%%%%%%%%%%%%%%%%%
%FIGURE 2
\begin{figure} [t] 
\includegraphics[scale=0.8]{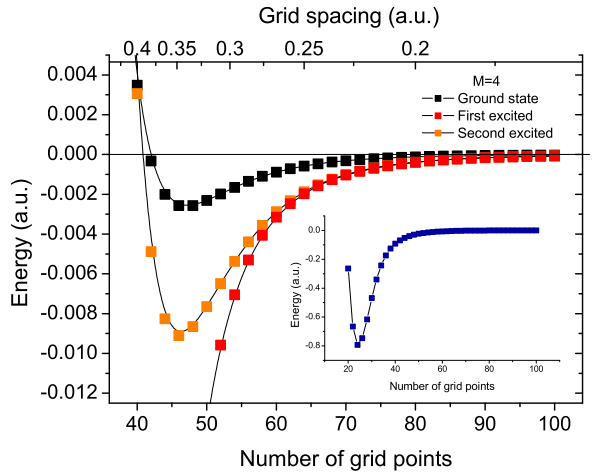}
\caption{(Color online) Convergence of the eigenvalues of the
one dimensional Schr\"{o}dinger equation discussed in the text
versus the number of grid points. The inset shows the eigenvalues sum.
}
\label{fig2}
\end{figure}
%%%%%%%%%%%%%%%%%%%%%%%%%%%%%
%%%%%%%%%%%%%%%%%%%%%%%%%%%%%
%FIGURE 3
\begin{figure} [t] 
\includegraphics[ scale=0.32]{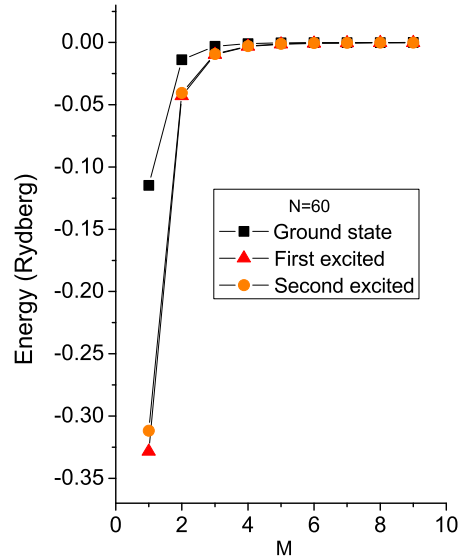}
\caption{(Color online) Convergence of the eigenvalues of the
one dimensional Schr\"{o}dinger equation discussed in the text
versus the representation order $M$.
}
\label{fig3}
\end{figure}
%%%%%%%%%%%%%%%%%%%%%%%%%%%%%
%FIGURE 4
\begin{figure} [t] 
\includegraphics[ scale=0.3]{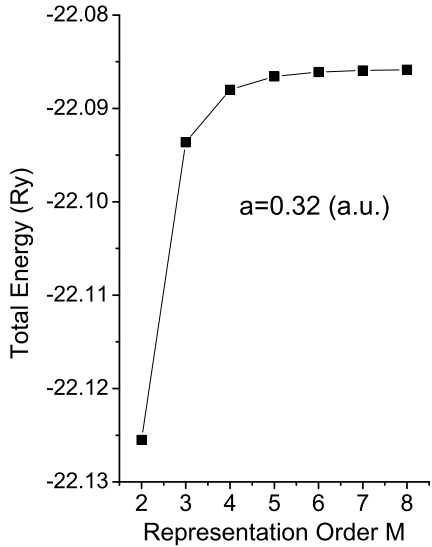}
\caption{(Color online) Convergence of the total energy of a C$_2$
molecule versus the representation order $M$ with a grid spacing
$a=0.32$ atomic units.
}
\label{fig4}
\end{figure}
%%%%%%%%%%%%%%%%%%%%%%%%%%%%%%
%FIGURE 5
\begin{figure} [t] 
\includegraphics[ scale=0.3]{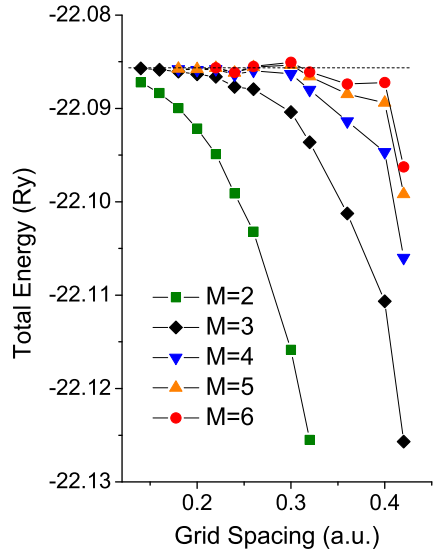}
\caption{(Color online) Convergence of the total energy of  a C$_2$
molecule versus the grid spacing for different values of 
the representation order $M$.}
\label{fig5}
\end{figure}
%FIGURE 5
%%%%%%%%%%%%%%%%%%%%%%%%%%%%%%%
\section{Numerical examples}
\label{sec5}

The convergence properties discussed in the previous section
are here verified
on two very different numerical examples. The first one is the 
one dimensional Schr\"{o}dinger equation with the potential
$V(x)=- U_0/cosh^2(\alpha x)$ for which an exact analytical
solution is available\cite{landau}. Choosing $U_0=-21.0 Ry$ and
$\alpha=1.4$, it can be seen that the bound states are just three.

Using the kinetic energy matrix elements of Eq. (\ref{cin5a}), we have
numerically calculated the three eigenvalues changing both $M$ and 
$N$. The convergence towards the exact eigenvalues on changing $N$ is shown
in Fig. \ref{fig2} while that obtained on changing $M$ is shown
in Fig. \ref{fig3}. Moreover, in the inset of Fig. \ref{fig2} we also show
the convergence of the eigenvalues sum. The data in Fig. \ref{fig2}
show that a monotonic convergence from below is 
recovered only for sufficiently large $N$. On the contrary, as shown in 
Fig. \ref{fig3}, the convergence with respect to $M$
is fully monotonic. The difference is due to
the way the potential contributes to the final result. In Fig. \ref{fig2}
we change the grid spacing so that the potential is sampled differently
for each $N$ whereas in Fig. \ref{fig3} we are only changing 
the representation order in the kinetic
energy matrix. The convergence from below shown in Fig. \ref{fig3} is
consistent with the data of Fig. \ref{fig1}.    

The second numerical example we have analyzed is that
of a DFT calculation of the ground state of a C$_2$ molecule.
The calculations have been performed
 using the finite difference pseudopotential method
developed by Chelikowsky et  al\cite{cheli1}. In Fig.\ref{fig4} we show the convergence
of the total energies with respect to the representation order $M$
while in Fig.\ref{fig5} we show the convergence versus the grid spacing. In both the cases
the convergence from below due to the kinetic energy is quite evident.
Despite the enormous difference in the computational complexity
with respect to the one dimensional case, the general trends are similar.
%%%%%%%%%%%%%%%%%%%%%%%%%%%%%%%%%

A last example is here reported to evaluate, in bulk silicon, the convergence properties of the total energy with respect to the grid spacing and the representation order. We used Octopus scientific code to perform DFT real space calculations\cite{octopus}. 
%%%%%%%%%%%%%%%%%%%%%%%%%%%%%%
%FIGURE 6
%\begin{figure} [t] 
%\includegraphics[ scale=0.3]{Figure6.eps}
%\caption{(Color online) Convergence of the total energy of  bulk silicon
 %versus the representation order, at a fixed grid spacing $a=0.4$ a.u.}
%\label{fig6}
%\end{figure}
%FIGURE 5
%%%%%%%%%%%%%%%%%%%%%%%%%%%%%%%
%%%%%%%%%%%%%%%%%%%%%%%%%%%%%%
%FIGURE 7
\begin{figure} [t] 
\includegraphics[ scale=0.3]{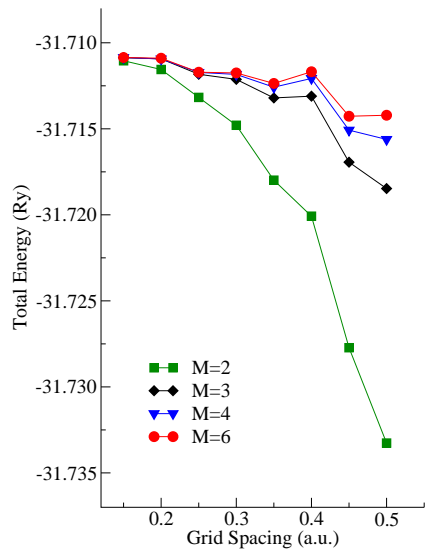}
\caption{(Color online) Convergence of the total energy of  bulk silicon
 versus the grid spacing, for several values of the representation order.}
\label{fig6}
\end{figure}
%FIGURE 5
%%%%%%%%%%%%%%%%%%%%%%%%%%%%%%%
A 2x2x2 Monkhorst Pack grid was adopted to perform  integrations over the first Brillouin zone, together with a Perdew-Zunger LDA exchange-correlation functional, and a 8-atom cubic conventional cell, with a lattice constant fixed at $10.3$ a.u.
The total energy was evaluated on changing the representation order and the grid spacing, each time redoing the selfconsistent procedure.
The convergence of the total energy with respect to the grid spacing, for several values of $M$, is reported in figure \ref{fig6}. The scaling is qualitatively similar to the  example illustrated in the previous section, and the results discussed above are confirmed. 
Even in this case, the convergence is from below. The total energy is fully monotonic with respect to  the representation order (at fixed grid spacing), and shows a striking similarity to figure \ref{fig4} (for this reason the graph is not reported here). Instead, as reported in figure \ref{fig6}, the convergence is always from below, but with oscillating features, when the grid spacing decreases at a fixed value of $M$. It is worth noticing that a higher representation order leads to a faster convergence with respect to the grid spacing. A tradeoff, in terms of computational time, has to be found between representation order and grid spacing. However, the use of a high value of $M$ is surely convenient. The default value chosen by the code for the representation order is $M=4$.

\section{Conclusions}
%%%%%%%%%%%%%%%%%%%%%%%%%%%%%%%
With this work we have made an attempt of resolving 
two main misconceptions
related to finite difference electronic structure calculations. The first is that
it is not really true that a finite difference representation has not a
basis set and the second is that the anti variational behavior is not a
limitation of the method. 

The proof that a basis set does exist has been a constructive one in the 
sense that we have derived explicit expressions for the momentum
and kinetic energy matrix elements showing that they are 
identical to the central finite difference weights of the first and
second derivatives.     

The convergence from below has been fully characterized
calculating the momentum and kinetic energy eigenvalues using
periodic boundary conditions. We have derived for both 
the operators two formulas, Eqs. (\ref{eq36}) and (\ref{eq39}), 
that we think can be useful for
developing convergence criteria. We hope to have made clear
that the anti variational behavior is an intrinsic
feature of finite difference method intimately connected to the fact
for each couple of the integer number $N$ and $M$ one has
a different representation of the operators. 

An important point of this work is that the continuum momentum and
kinetic energies can be obtained
with two different limits. 
The first one is obtained taking the limit of large
values of $M$, the representation order, with a fixed grid spacing. 
This limit, taken on a finite grid, has a meaning only when periodic boundary conditions
are imposed. The exact linear momentum and kinetic energy of a free
particle on a lattice are recovered. 
The second limit  consists in sending the grid spacing to zero
while keeping $M$ unchanged. We have shown that also
in this case the continuum momentum and kinetic energy
are recovered.

A concluding remark may be that the above result 
may be useful for reviewing,  
from a different perspective, the quantum mechanics
in a discretized space.

\newpage
%\section*{Acknowledgements}
%\section*{References}
\providecommand{\latin}[1]{#1}
\providecommand*\mcitethebibliography{\thebibliography}
\csname @ifundefined\endcsname{endmcitethebibliography}
  {\let\endmcitethebibliography\endthebibliography}{}

\end{document}